\documentclass[sigconf, nonacm, screen]{acmart}
\usepackage{listings}
\usepackage{makecell}
\usepackage{multirow}
\usepackage{multicol}
\usepackage{pifont}

\AtBeginDocument{%
  }

\begin{document}

\title[MM-CTR: Multimodal CTR Prediction Challenge]{1\textsuperscript{st} Place Solution of WWW 2025 EReL@MIR Workshop Multimodal CTR Prediction Challenge}

\author{Junwei Xu}
\affiliation{%
  \department{Shenzhen Key Laboratory of Ubiquitous Data Enabling, Shenzhen International Graduate School}
  \institution{Tsinghua University}
  \city{Shenzhen}
  \country{China}
}
\email{xjw23@mails.tsinghua.edu.cn}

\author{Zehao Zhao}
\affiliation{%
  \department{Shenzhen International Graduate School}
  \institution{Tsinghua University}
  \city{Shenzhen}
  \country{China}
}
\email{zhaozeha23@mails.tsinghua.edu.cn}

\author{Xiaoyu Hu}
\affiliation{%
  \department{Shenzhen International Graduate School}
  \institution{Tsinghua University}
  \city{Shenzhen}
  \country{China}
}
\email{huxiaoyu23@mails.tsinghua.edu.cn}

\author{Zhenjie Song}
\affiliation{%
  \department{Shenzhen International Graduate School}
  \institution{Tsinghua University}
  \city{Shenzhen}
  \country{China}
}
\email{songzj23@mails.tsinghua.edu.cn}

\renewcommand{\shortauthors}{Junwei Xu et al.}

\begin{abstract}
  The WWW 2025 EReL@MIR Workshop Multimodal CTR Prediction Challenge\footnote{\url{https://erel-mir.github.io/challenge}} focuses on effectively applying multimodal embedding features to improve click-through rate (CTR) prediction in recommender systems. 
  This technical report presents our 1\textsuperscript{st} place winning solution for Task 2, combining sequential modeling and feature interaction learning to effectively capture user-item interactions.
  For multimodal information integration, we simply append the frozen multimodal embeddings to each item embedding.
  Experiments on the challenge dataset demonstrate the effectiveness of our method, achieving superior performance with a 0.9839 AUC on the leaderboard, much higher than the baseline model.
  Code and configuration are available in our GitHub repository\footnote{\url{https://github.com/pinskyrobin/WWW2025_MMCTR}} and the checkpoint of our model can be found in HuggingFace\footnote{\url{https://huggingface.co/pinskyrobin/WWW2025_MMCTR_momo}}.
\end{abstract}



\maketitle

\section{Introduction}
Click-through rate (CTR) prediction is a fundamental task in online advertising and recommendation systems, aiming to estimate the probability of users clicking on specific ads or content items. 
The accuracy of CTR prediction directly affects the revenue of advertising platforms and the user experience. 
Traditional CTR models predominantly rely on structured tabular data (\emph{e.g.}, user demographics, historical behaviors, and item attributes), using feature interaction paradigms like factorization machines (FMs) \cite{FM, FFM, DeepFM, NFM, xDeepFM} or advanced deep learning techniques \cite{DIN, DCNv2, SASRec, AutoInt, TransAct}.
However, the exponential growth of multimodal content (\emph{e.g.}, video covers, content title and audio clips) has become increasingly available in recommender systems, necessitating the development of advanced methods that can effectively leverage the multimodal information.

The Multimodal CTR Prediction (MM-CTR) Challenge at the WWW 2025 EReL@MIR Workshop\cite{EReL} aims to foster innovation in adopting multimodal embedding features for CTR prediction.
In response to industry requirements for low-latency and online inference, the challenge comprises two complementary sub-tasks: Multimodal Item Representation Learning and Multimodal CTR Modeling. 
The first sub-task focuses on learning multimodal item representations optimized for recommendation scenarios, while the second emphasizes effectively leveraging these frozen multimodal embeddings to further enhance the performance of the CTR model. 
We participated in the second sub-task, which can be divided into two parts: better utilization of frozen multi-modal embedding features and better CTR prediction modeling.

Although injecting multimodal semantic information into the CTR model was explored, simple concatenation of multimodal embeddings with item embeddings is adopted in our final solution. 
This was due to the limited time, and we were unable to finish the model optimization and parameter tuning work of the multimodal embeddings part.
Inspired by \cite{TransAct}, we adopt Transformer for sequential modeling and DCNv2\cite{DCNv2} for feature interaction learning.
Through extensive parameter tuning, the optimal hyperparameter settings were obtained on the challenge dataset.
We achieved the 1\textsuperscript{st} place on the final leaderboard with an AUC score of 0.9839, demonstrating the effectiveness of our method.

\section{Methods}

\subsection{Problem Formulation}
Given a set of samples $\mathcal{D} = \{(\mathcal{H}_u, x_{target}, y) | u\in\mathcal{U}, x_{target}\in\mathcal{I}, y\in\{0, 1\})\}$, 
where \(\mathcal{U}\) and \(\mathcal{I}\) are the user set and item set. $\mathcal{H}_u = \{x_1, x_2, \ldots, x_N\}$ is the historical interaction sequence of user \(u\), and $x_i$ is the item features (\emph{e.g.}, item ID, item tags and item multimodal embedding) of the \(i\)-th clicked item in the user history. 
\(y\) is a binary label indicating whether the user clicked on the target item or not:
\begin{equation}
  y = f\left(\mathcal{H}_u, x_{target}| \mathcal{D}, \Theta\right),
\end{equation}
where $\Theta$ is the model parameter.

\subsection{Network Structure}
\begin{figure}[htbp]
  \vspace{-1mm}
  \centering
  \includegraphics[width=\linewidth]{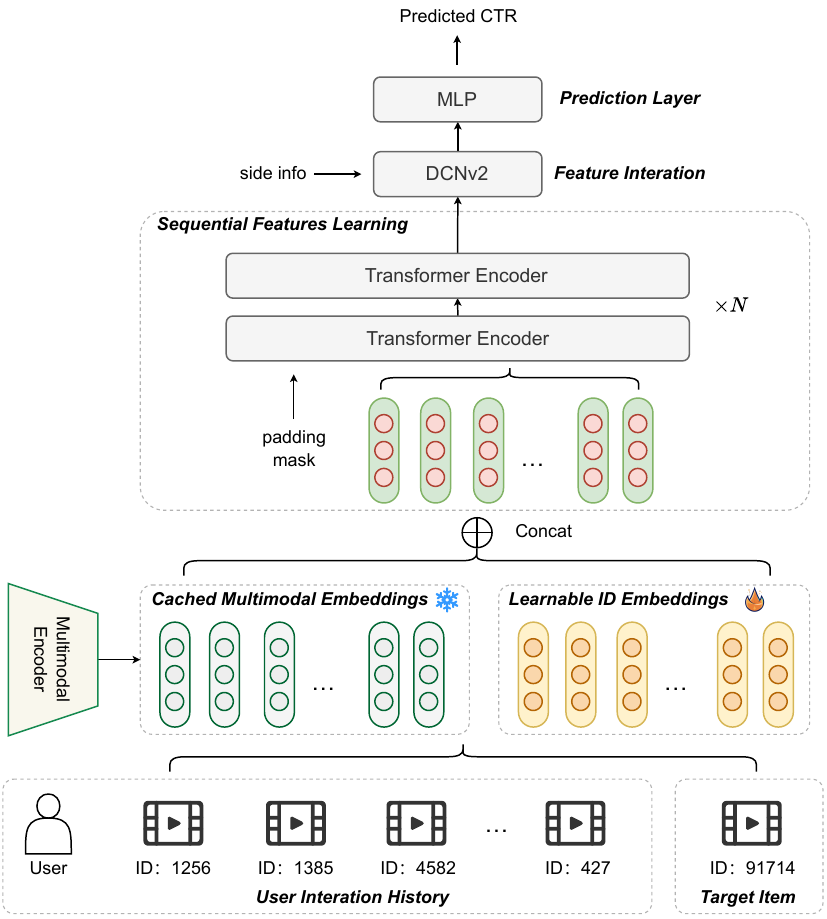}
  \caption{The overall architecture of our model.}
  \label{fig:network}
  \vspace{-2mm}
\end{figure}

Inspired by \cite{TransAct}, sequential modeling and feature interaction learning are combined to effectively capture user interest preferences.
As shown in Figure \ref{fig:network}, it consists of four main components: Embedding Layer, Sequential Feature Learning Module, Feature Interaction Module, and Prediction Layer.

\subsubsection{Embedding Layer}
The Embedding Layer converts different types of input features into dense vector representations.
Let \(x=\{t_1, t_2,\ldots,t_{|T|},e_{mm}\}\) be the item features, where \(t_i\) is the \(i\)-th feature of the item, \(|T|\) is the number of features and \(e_{mm}\) is the frozen multimodal embedding of the item.
The embedding layer maps each feature \(t_i\) to a dense vector \(e_i\) using an embedding matrix:
\begin{equation}
  e_{t_i} = E(t_i) \in \mathbb{R}^{d_e},
\end{equation}
where \(d_e\) is the embedding dimension.
The frozen multimodal embedding \(e_{mm}\) is then concatenated with the item embeddings to form the final item representation:
\begin{equation}
  e^i_{item} = \left[e^i_{t_1} \parallel e^i_{t_2} \parallel \ldots \parallel e^i_{t_{|T|}} \parallel e^i_{mm}\right],
\end{equation}
where \(\parallel\) denotes the concatenation operation.

\subsubsection{Sequential Feature Learning Module}
We use Transformer to capture temporal patterns.
The target item embedding \(e_{target}\) is concatenated with every item embedding in the history sequence to form the input sequence:
\begin{equation}
  \widetilde{e^i}_{item} =\left[e^i_{item} \parallel e_{target}\right],
\end{equation}
where \(e_{target}\) is the embedding of the target item.

The length of the history sequence is \(N\). And for those users with fewer than \(N\) interactions, the sequence will be padded with zeros.
The input sequence is then fed into a Transformer layer with several Transformer Encoders:
\begin{equation}
  \begin{aligned}
    S & = \left(s_1,s_2,\ldots,s_N\right) \\ & = \texttt{Transformer}\left(\left[\widetilde{e^1}_{item},\widetilde{e^2}_{item},\ldots,\widetilde{e^N}_{item}\right]\right).
  \end{aligned}
\end{equation}

The output of the Transformer layer is \(S\in\mathbb{R}^{N\times d_t}\), where \(d_t\) is the dimension of the Transformer output for each item.
While the user's interest cannot be fully represented by the last output, directly using all the outputs of all the items in the history sequence will significantly increase the complexity of the model.
Following \cite{TransAct}, the latest \(k\) outputs are selected as the representation of user's short-term interest preference.
And the max pooling operation is adopted to represent the user's long-term interest preference:
\begin{equation}
    \label{eq:seq}
    S_{o} = \texttt{Flatten}\left(s_1,s_2,\ldots,s_k,\texttt{MaxPool}\left(S\right)\right),
\end{equation}
where \texttt{Flatten} is the flattening operator along the last dimension and \texttt{MaxPool} is the max pooling operator.

\subsubsection{Feature Interaction Module}
To explicitly model the interactions between features, we adopt DCNv2\cite{DCNv2} as the feature interaction module for its efficiency and effectiveness of modeling high-order feature interactions:
\begin{equation}
  \begin{gathered}
  f_i = \left[e_{target}, e_{side}, S_o \right], \\
  c_{l+1} = f_i\odot \left(W_l c_l+b_l\right) + c_l, \\
  d_o = \texttt{MLP}_f\left(f_i\right),
  \end{gathered}
\end{equation}
where \(e_{side}\) is the concatenated embedding of the side features (\emph{e.g.}, like level and view level of the target item). 
\(c_l\), \(W_l\) and \(b_l\) are the output, weight and bias of the \(l\)-th cross layer.
\(\odot\) denotes the element-wise multiplication.
\texttt{MLP}\(_l\) is a 3-layer perceptron with ReLU activation function, representing the deep network part of DCNv2.
The parallel structure is adopted in our DCNv2 module, thus the output of the feature interaction module is:
\begin{equation}
  f_o =\left[c_o, d_o\right],
\end{equation}
where \(c_o\) is the final output of the last cross layer.

\subsubsection{Prediction Layer}
Since CTR prediction is a binary classification task, we use a 2-layer perceptron and a sigmoid function to predict the probability of the target item being clicked:
\begin{equation}
  \hat{y} = \sigma\left(\texttt{MLP}_p\left(f_o\right)\right),
\end{equation}
where \(\sigma\) is the sigmoid function.

\subsubsection{Loss Function}
The loss function is defined as the binary cross-entropy loss:
\begin{equation}
  \mathcal{L} = -\frac{1}{|\mathcal{D}_{tr}|}\sum_{i=1}^{|\mathcal{D}_{tr}|}\left[y_i\log(\hat{y}_i) + (1-y_i)\log(1-\hat{y}_i)\right],
\end{equation}
where \(|\mathcal{D}_{tr}|\) is the number of samples in the training set.

\section{Experiments}

\subsection{Preparations}
\subsubsection{Datasets}
The dataset\footnote{\url{https://recsys.westlake.edu.cn/MicroLens_1M_MMCTR/}} provided by the MM-CTR Challenge originates from the recently released MicroLens dataset by Westlake University\cite{MicroLens}. 
It contains 1M users and 91.7K items, with each item featuring rich modalities including text descriptions, images, audio, and raw video information.
To obtain the multimodal embeddings, we use the PCA embedding of the concatenated BERT\cite{BERT} and CLIP\cite{CLIP} embeddings as the multimodal embedding.
The whole dataset is divided into training, validation, and test sets, with 3.6M, 10k, and 380k samples respectively.

\subsubsection{Environmental Setup}
We have released our code and configuration files on GitHub\footnote{\url{https://github.com/pinskyrobin/WWW2025_MMCTR}}.
All of our code is implemented based on FuxiCTR\footnote{\url{https://github.com/reczoo/FuxiCTR}}.
After cloning the repository, the environment can be replicated using the following commands:

\begin{lstlisting}[
  language=python,
  basicstyle=\ttfamily,
  breaklines=true,
  keywordstyle=\bfseries\color{blue},
  morekeywords={},
  emph={self},
     emphstyle=\bfseries\color{Rhodamine},
     commentstyle=\itshape\color{black!50!white},
     stringstyle=\bfseries\color{PineGreen!90!black},
     columns=flexible,
    %  numbers=left, % 显示行号
    %  numbersep=2em, % 设置行号的具体位置
    %  numberstyle=\footnotesize % 缩小行号
     frame=single
 ]  
  conda create -n fuxictr_momo python==3.9
  pip install -r requirements.txt
  source activate fuxictr_momo
\end{lstlisting}

All experiments were conducted on a customized GPU with 32GB VRAM (vGPU-32G) through AutoDL\footnote{\url{https://www.autodl.com}}. 
The versions of CUDA and PyTorch are 11.7 and 1.13.1 respectively.

\subsubsection{Parameter Settings}
\begin{table}[t!]
  \centering
  \caption{Hyperparameters of our model.}
  \label{tab:hyperparameters}
  \resizebox{\linewidth}{!}{
    \begin{tabular}{ccc}
      \toprule
      Hyperparameter & Grids & Best Value \\
      \midrule
      \texttt{learning\_rate} & [1e-3, 5e-4, 5e-5, 1e-5] & 5e-4 \\
      \texttt{embedding\_dim} & [16, 32, 64, 128] & 64 \\
      \texttt{transformer\_dropout} & [0, 0.1, 0.2, 0.3, 0.4] & 0.2 \\
      \texttt{cross\_net\_dropout} & [0, 0.1, 0.2, 0.3, 0.4] & 0.2 \\
      \texttt{k} in Eq.\ref{eq:seq} & [0, 2, 4, 8, 16, 24] & 16 \\
      \bottomrule
    \end{tabular}
  }
\end{table}
We use the standard Adam optimizer with a learning rate of \(5e^{-4}\) and a batch size of 128.
Embedding dimension is set to 64. Numbers of cross layers and Transformer encoders are set to 3 and 2 respectively.
Hidden units of the deep network in DCNv2 and the prediction layer are set to \([1024, 512, 256]\) and \([64, 32]\) respectively.
Dropout rate in both Transformer and DCNv2 is set to 0.2.
\(k\) in Eq.\ref{eq:seq} is set to 16.
Early stopping is applied to avoid model overfitting.
The training process will be terminated if the validation AUC score does not improve for 5 consecutive epochs.
We carefully tuned specific hyperparameters in our model by grid search, detailed in Table \ref{tab:hyperparameters}.

\subsubsection{Evaluation Metrics}
Our model is evaluated using the area under the ROC curve (AUC) and the log loss metrics.
Both metrics are widely used in CTR prediction tasks.
AUC measures the model's ability to distinguish between positive and negative samples, while log loss quantifies the model's prediction accuracy.
The higher the AUC and the lower the log loss, the better the model's performance.

\subsection{Overall Performance}
\begin{table}
  \centering
  \caption{Overall performance evaluation.}
  \label{tab:results}
  \resizebox{\linewidth}{!}{
    \begin{tabular}{cccc}
      \toprule
      Model & \emph{w/ multimodal emb.} & AUC & Logloss \\
      \midrule
      \multirow{2}{*}{Baseline (DIN)} & \ding{55} & 0.9326 & 0.6485 \\
      & \ding{51} & 0.8577 & 2.7697 \\
      \cmidrule(lr){2-4}
      \multirow{2}{*}{\emph{w/ Dice}} & \ding{55} & 0.9366 & 0.6878 \\
      & \ding{51} & 0.8829 & 2.5390 \\
      \midrule
      \multirow{2}{*}{Ours} & \ding{55} & 0.9729 & 0.2369 \\
      & \ding{51} & \textbf{0.9776} & \textbf{0.2358} \\
      \cmidrule(lr){2-4}
      \multirow{2}{*}{\emph{w/o Transformer}} & \ding{55} & 0.9741 & 0.2617 \\
      & \ding{51} & 0.9688 & 0.3379 \\
      \cmidrule(lr){2-4}
      \multirow{2}{*}{\emph{w/o DCNv2}} & \ding{55} & 0.9023 & 0.4996 \\
      & \ding{51} & 0.9632 & 0.3522 \\
      \bottomrule
    \end{tabular}
  }
  \vspace{-2mm}
\end{table}

The baseline model provided by the challenge organizers is a DIN model \cite{DIN} without Dice activation function.
The overall performance comparison of different models with and without multimodal embeddings is summarized in Table~\ref{tab:results}. 
AUC improvement is witnessed when we simply replace the activation function with Dice.
Our model achieves the best performance across both evaluation metrics, demonstrating its superiority over the baseline.

It is worth noting that the baseline DIN model and our model without Transformer suffer a substantial performance degradation when multimodal embeddings are added, suggesting limited compatibility with multimodal features.
We suspect that the frozen multimodal embeddings are not well aligned with the CTR prediction task, making it difficult for the model optimization and better utilization of multimodal information.

Ablation studies reveal the importance of the two key components.
Removing the DCNv2 layer (\emph{w/o DCNv2}) sharply degrades performance, while omitting the Transformer module (\emph{w/o Transformer}) also leads to a noticeable decrease in AUC and an increase in log loss, highlighting the complementary roles of sequential feature learning and cross-feature interaction. 
Overall, the results confirm that our full model optimally integrates multimodal embeddings with Transformer and DCNv2 components to maximize predictive accuracy.
And an AUC of 0.9839 on the leaderboard was achieved, ranking 1\textsuperscript{st} in the challenge.

\subsection{Parameter Sensitivity Analysis}
\begin{figure*}[t!]
  \centering
  \includegraphics[width=0.82\linewidth]{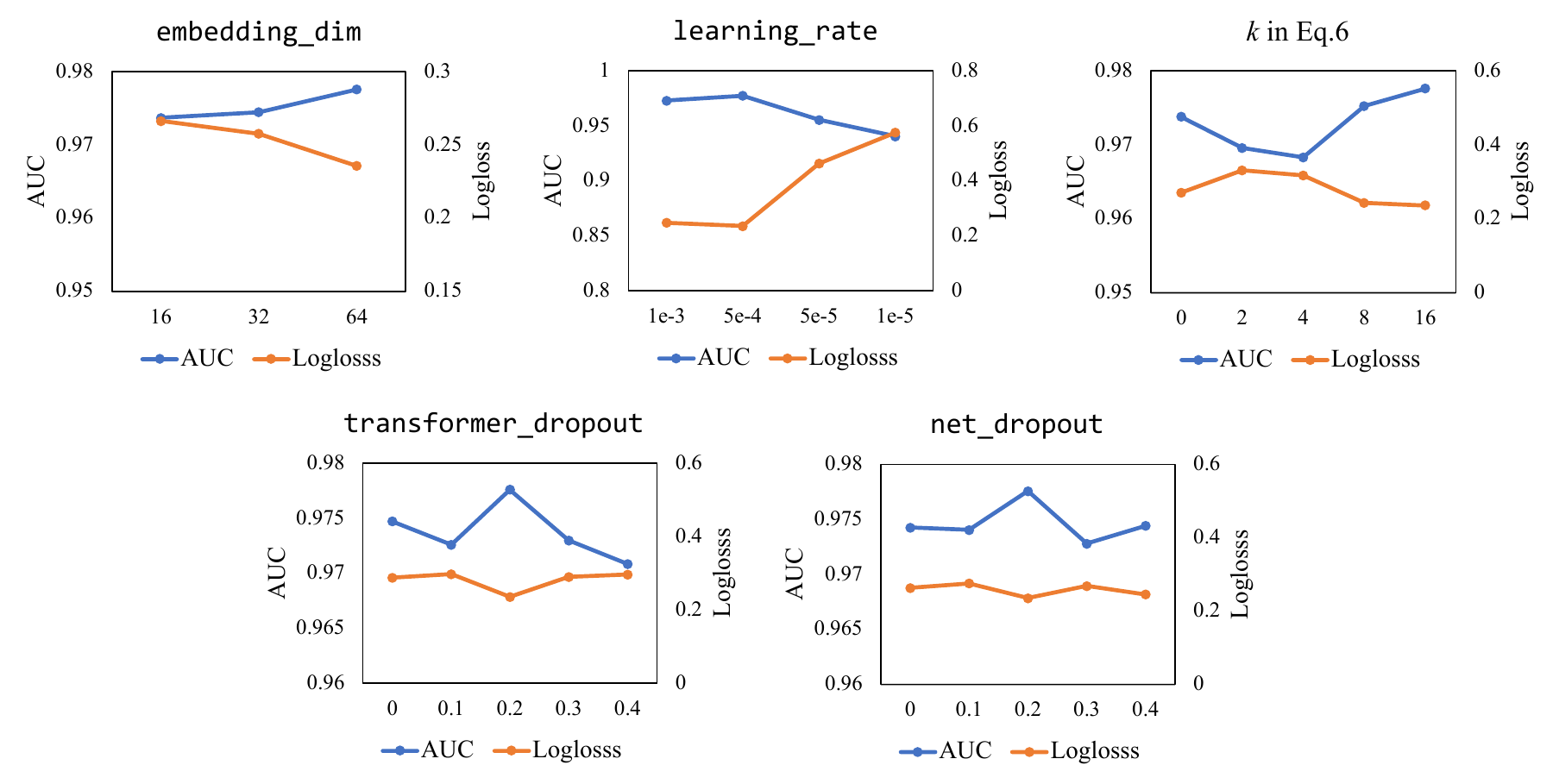}
  \caption{Parameter sensitivity analysis on hyperparameters.}
  \label{fig:ps}
\end{figure*}

We conduct parameter sensitivity analysis on various hyperparameters, as shown in Figure \ref{fig:ps}.
Assigning specific values to certain hyperparameters may cause the model collapse phenomenon (\emph{e.g.}, high learning rate), and will not be reported.

Experiments show that the model is sensitive to the learning rate, and a learning rate of \(5e^{-4}\) works significantly better than other values.
\texttt{embedding\_dim} and \texttt{k} in Eq.\ref{eq:seq} significantly affect the model capacity and training efficiency, while contributing less to the final performance.
Appropriate \texttt{dropout} settings have a positive but limited effect on the model performance, whose adjustments were placed at the end to push the performance to the limit.

\section{Conclusion}
Although model ensemble is not allowed in the challenge, the superior performance could still be presented by the simple yet effective model architecture.
FuxiCTR provides efficient configuration for tuning our model, saving us a lot of time.
Due to the limited time, we simply concatenate the multimodal embeddings with item embeddings.
Aligning multi-modal embeddings with the downstream CTR task is the key to further improving the performance of the model.
In the future, we will explore quantization methods to transform the frozen multimodal embeddings into semantic and learnable item embeddings.
How to effectively utilize prior knowledge of the multimodal embeddings to guide the learning of collaborative ID embeddings is also a worthy research direction.


\bibliographystyle{ACM-Reference-Format}
\bibliography{main}

\end{document}